\documentclass[preprint,12pt]{elsarticle}
\usepackage{amssymb}
\usepackage{epsfig}
\journal{Journal of Magnetism and Magnetic Materials}
\begin{document}
\begin{frontmatter}
\title{Tailoring the ground state of the ferrimagnet
{La$_{2}$Ni(Ni$_{1/3}$Sb$_{2/3}$)O$_6$}}

\author[bari,cord]{D. G. Franco}
 \address[bari]{Laboratorio de Bajas Temperaturas, Centro At\'omico Bariloche - CNEA, 8400 Bariloche, R\'io Negro,
Argentina.}
\address[cord]{INFIQC-CONICET, Depto. de F\'{i}sico Qu\'{i}mica,
Facultad de Ciencias Qu\'{i}micas, Universidad Nacional de
C\'ordoba, Ciudad Universitaria, X5000HUA C\'ordoba, Argentina.}
\author[cord]{R. E. Carbonio}
\author[bari,bariib]{E. E. Kaul}
\author[bari,bariib]{G. Nieva}\ead{gnieva@cab.cnea.gov.ar}
\address[bariib]{Instituto Balseiro, CNEA and Universidad Nacional de Cuyo,
8400 Bariloche, R\'io Negro, Argentina.}

\begin{abstract}
We report on the magnetic and structural properties of
La$_{2}$Ni(Ni$_{1/3}$Sb$_{2/3}$)O$_6$ in polycrystal, single
crystal and thin film samples. We found that this material is a
ferrimagnet ($T_{c}\!\approx$ 100 K) which possesses a very
distinctive and uncommon feature in its virgin curve of the
hysteresis loops. We observe that bellow 20 K it lies outside the
hysteresis cycle, and this feature was found to be an
indication of a microscopically irreversible process possibly
involving the interplay of competing antiferromagnetic
interactions that hinder the initial movement of domain walls.
This initial magnetic state is overcome by applying a temperature
dependent characteristic field. Above this field, an isothermal
magnetic demagnetization of the samples yield a  ground state different
from the initial thermally demagnetized one.

\end{abstract}

\begin{keyword}
Ferrimagnetics \sep Doubleperovskites \sep Magnetic frustration \sep Superexchange and super-superexchange interactions \sep Magnetic oxides

\end{keyword}
\end{frontmatter}

\section{Introduction}
The physics and chemistry of complex oxides have been largely
studied due to the rich phenomena resulting from their combined
magnetic, charge and orbital degrees of freedom. Among these
complex oxides, those with perovskite structure  ABO$_{3}$ have
been extensively studied. They are the archetype of
superconducting oxides, giant magnetoresistive oxides and the
materials for the emergent field of novel device functions
\cite{Takagi}.

The size and oxidation state of A and B cations determine the
symmetry of the perovskite structure, that departs from cubic when
a tilting of the octahedral arrangement of oxygen around the B
cation occurs \cite{Levin}. Also, the partial replacement of A or B
site cations by isovalent or aliovalent A' and B' atoms could
result in a double perovskite structure with lower symmetry. In
double perovskites with general formula A$_{2}$(BB')O$_{6}$, B and
B' atoms occupy two different crystallographic sites, and an
ordered or disordered occupancy of them depends on the oxidation
state and size difference between B and B' ions \cite{Kobayashi,
Holman}.

Many of the magnetic interactions found in transition metal oxide
perovskites are due to superexchange and/or super-superexchange
interactions mediated through the oxygen orbitals. In some
materials the relative strength and arrangement of these interactions
determine the magnetic structure, range of the ordering temperatures
and the possibility of frustration \cite{Koo, Viola, Maignan, Battle,
Murthy}. In particular, the effective spin lattice, ferro- or
antiferromagnetic interactions among transition metal ions, depends
on a delicate balance provided by the charge in the linking
orbital, the bond angle, the degree of orbital overlap, the
distance between interacting ions and their spin state
\cite{Goodenough, Hoffman}. In the specific case of the perovskite
structure, the typical bond angles and distances usually favors
antiferromagnetic superexchange interactions \cite{Hoffman}.
However, in some particular cases, when there are more than one
electronic pathway in the linking orbitals, a destructive
interference leads to the cancellation of the antiferromagnetic
interaction. In these cases, a weak ferromagnetic coupling becomes
relevant \cite{Levstein}.

In transition metal oxides, the competing effects of a
ferromagnetic interaction and local frustration leads to a spin
glass like behavior \cite{Serrate}. The main signatures of this
frustration are made evident in the time evolution from metastable
magnetic states \cite{Li}. However, in bulk ferro/ferrimagnetic
samples, a local frustration is normally hard to visualize due to
the magnetic history dependence of the metastable states created
by the pining of domain walls.

Also in some transition metal oxides the competition between a
ferro- and antiferromagnetic ground states leads to a
nonequilibrium glassy behavior that arises from a kinetic arrest
of a ferromagnetic to antiferromagnetic phase transition
\cite{SarkarII}. These materials  are identified as magnetic
glasses \cite{Roy} consisting of ferro- (or ferri-) magnetic and
antiferromagnetic clusters frozen randomly with a dynamics similar
to that of structural glasses.

This article will present a detailed magnetic study of the
ferrimagnetic double perovskite
La$_{2}$Ni(Ni$_{1/3}$Sb$_{2/3}$)O$_6$ \cite{Alvarez, AlvarezII}
showing evidence of a frustrated magnetic state at low temperatures. The
properties of this compound were explored using polycrystalline, single
crystalline and thin film samples. The main result is the observation that
below 20 K there is evidence of a microscopically irreversible process
involving the interplay of competing antiferromagnetic
interactions that hinders the initial magnetic polarization or the
movement of domain walls and determines the microscopic nature of
the strong pinning centers found in this system \cite{IEEE Arxiv}.
This material results in a model system for studying the seldom found
ferrimagnetic Ni$^{2+}$ oxides and the low temperature glassy or
magnetically disordered state present in many complex transition
metal oxides.

\section{Experimental details}

We prepared polycrystalline samples of
La$_{2}$Ni(Ni$_{1/3}$Sb$_{2/3}$)O$_6$ by conventional solid-state
reaction. Stoichiometric amounts of La$_2$O$_3$,
Ni(NO$_3$)$_2$.6H$_2$O and Sb$_2$O$_3$ were ground and fired at
1400$^{o}$C for 12 hours in air.  The single crystals were grown
by the floating zone technique in a double ellipsoidal mirror
furnace. The thin films were grown using RF magnetron sputtering
in an Ar/O$_2$ atmosphere on (100) SrTiO$_3$ substrates at
700-800$^{o}$C. We checked the composition of all the investigated
samples by EDS. Only the thin films showed a 10$\%$ Ni$^{2+}$ deficiency
not present in the stoichiometric target material. We carried out
the structural analysis using powder X-ray diffraction (XRD) at
room temperature. The magnetic measurements were performed in an
QD-MPMS SQUID magnetometer in the range 2 to 300 K and -5 to 5 Tesla.

\section{Results}
\subsection{Structural characterization}
\begin{figure}[h]
\includegraphics
[width=1\columnwidth, angle=0] {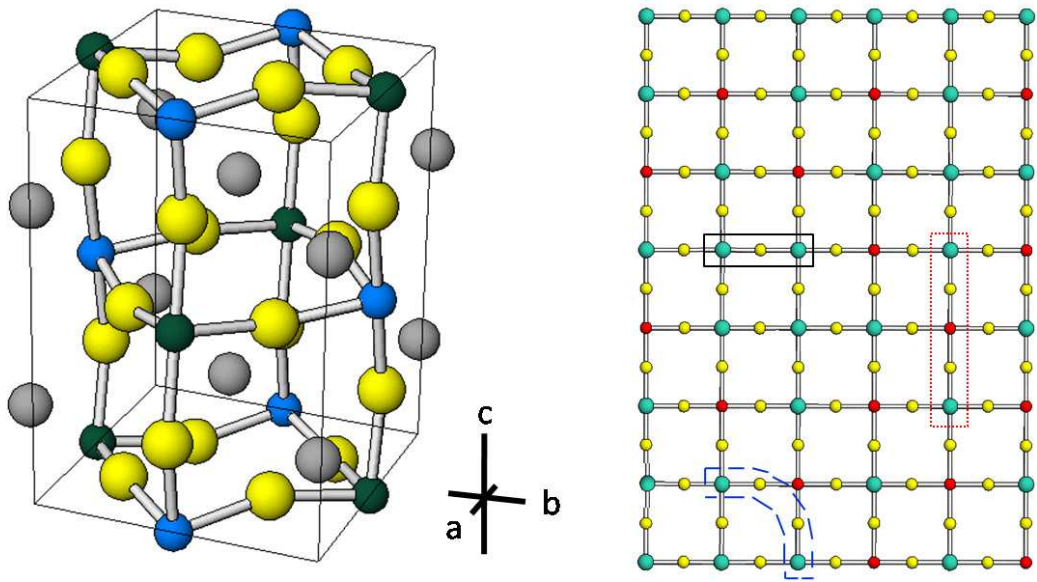} \caption[]{(color
online) Left: monoclinic structure of
La$_{2}$Ni(Ni$_{1/3}$Sb$_{2/3}$)O$_6$ double perovskite. Grey
spheres: La$^{3+}$, yellow spheres: O$^{2-}$, light blue spheres:
B$_{2d}$ ions and green spheres: B$_{2c}$ ions. Righ: two
dimensional scheme of the distribution of B ions over the 2d and
2c sites for La$_{2}$(Ni)$_{2d}$(Ni$_{1/3}$Sb$_{2/3}$)$_{2c}$O$_6$
showing the different neighbors of the magnetic Ni$^{2+}$ ions
(turquoise spheres): first next nearest neighbors (1 nnn, black),
second next nearest neighbors (2 nnn, dash blue line) and third
next nearest neighbors (3 nnn, red dot line). Red spheres are
Sb$^{5+}$ ions and yellow spheres O$^{2-}$ ions. For simplicity a
square structure was supposed and lanthanum ions were omitted.}
\label{structure}
\end{figure}

In spite of an earlier report of this structure as being
orthorrombic \cite{Alvarez}, with a fully disordered arrangement
of Ni$^{2+}$ and Sb$^{5+}$ ions, we found the crystalline symmetry
to be monoclinic (space group P2$_1/$n) with a rock salt
arrangement of BO$_6$ and B'O$_6$ octahedra described by the
a$^-$b$^-$c$^+$ system of three octahedral tilts in the Glazer's
notation.  The (Ni$^{2+}$/Sb$^{5+}$)$_{2d}$O$_6$ and
(Ni$^{2+}$/Sb$^{5+}$)$_{2c}$O$_6$
octahedra are rotated in phase along the primitive $c$ axis and
out-of phase along the primitive $a$ and $b$ axes. We performed a
Rietveld refinement of the structure using the FULLPROF program
\cite{Rodriguez-Carvajal}, obtaining the lattice parameters $a$ =
5.6051(3) \AA, $b$ = 5.6362(3) \AA, $c$ = 7.9350(5) \AA
\hspace{.05in} and $\beta$ = 89.986(4)$^o$. The occupancy of the
two crystallographic sites $2d$ and $2c$ were refined allowing the
Ni$^{2+}$/Sb$^{5+}$ distribution to vary, in order to model the
octahedral site disorder. We found that the $2d$ cation site is
almost fully occupied by Ni$^{2+}$ while the $2c$ site has an
occupancy close to 1/3 of Ni$^{2+}$ ions and 2/3 of Sb$^{5+}$.
The resulting crystallographic formula can be written as
La$_2$(Ni$_{0.976}$Sb$_{0.024}$)$_{2d}$(Ni$_{0.357}$Sb$_{0.643}$)$_{2c}$O$_6$.
It should be noted here that this space group does not allow a
further ordering of the Ni$^{2+}$ and Sb$^{5+}$ ions at $2c$ site.
Figure \ref{structure} shows the structure of
La$_{2}$Ni(Ni$_{1/3}$Sb$_{2/3}$)O$_6$ and also a schematic two
dimensional square view of Ni$^{2+}$/Sb$^{5+}$ distribution among
2d and 2c sites. From this picture it can be seen that Ni$^{2+}$
ions have three types of Ni$^{2+}$ neighbors: first next nearest
neighbors (mediated by -O-, that is, superexchange), second next
nearest neighbors (through a -O-O- bridge) and third next nearest
neighbors (-O-Sb$^{5+}$-O-, super-superexchange).

\begin{figure}[hhhhhhhhhhhhh]
\includegraphics[width=.9\columnwidth]{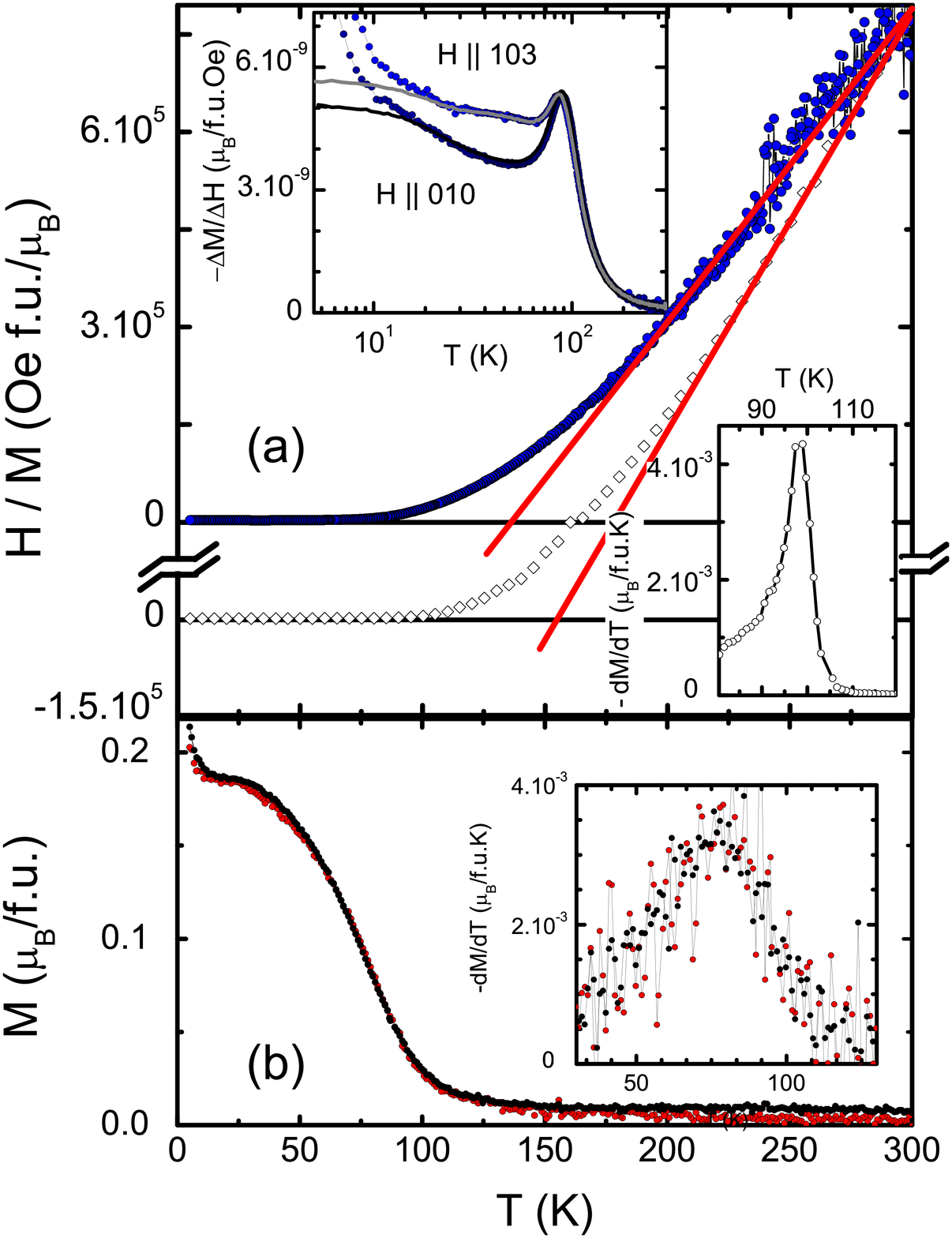} \caption[]{(color online)
(a) $H/M$ vs $T$ for a single crystalline (solid symbols) and a
polycrystalline (lower curve, open symbols)
La$_{2}$Ni(Ni$_{1/3}$Sb$_{2/3}$)O$_6$ samples, measured at 1000
Oe. The upper inset shows the susceptibility (defined in the
text), $\Delta$ $M$/$\Delta$ $H$ vs $T$ for $H$ along (103) and
(010) directions in the single crystal. The symbols indicate the
ZFC and the line the FC measurement.  The lower inset shows the
magnetization derivative $dM/dT$ vs $T$ for a polycrystalline
sample taken from a FC measurement at 1 Oe.  (b)$M$ vs $T$ for two
thin films of La$_{2}$Ni(Ni$_{1/3}$Sb$_{2/3}$)O$_6$, measured at
1000 Oe. The inset shows $dM/dT$ vs $T$ for the same films  FC
measurement at 1000 Oe.} \label{MvsT poly Xtal and film}
\end{figure}

Small crystals with a face parallel to the (103) planes could be
extracted from the rod grown in the mirror furnace. The rocking
curve on the (103) peak had a FWHM = 0.25$^o$ and the interplanar
spacing showed a 0.05$\%$ reduction with respect to the bulk.

The thin films (100 -130 nm thick) were grown epitaxially in the
$c$-axis direction. From XRD we measured a 1.4$\%$ $c$-axis
expansion as compared with bulk samples, considering a constant
cell volume. This corresponds to a  1.5$\%$
(Ni$^{2+}$/Sb$^{5+}$)$_{2d}$-(Ni$^{2+}$/Sb$^{5+}$)$_{2c}$ distance
reduction in the $ab$-plane, to match the Ti-Ti distance of the
substrate. Rocking curves showed typically a FWHM =0.4$^o$ indicating
a good crystallographic quality of the
La$_{2}$Ni(Ni$_{1/3}$Sb$_{2/3}$)O$_6$ films.

\subsection {Magnetic characterization}

We measured the magnetization as a function of temperature ($M$ vs
$T$) using a ZFC-FC procedure (i.e. cooling with zero applied
field or with a finite applied field) for several powder samples,
thin films and single crystals. The lower curve in Figure 2(a)
(open diamonds) shows a typical result for polycrystalline samples with a
Curie-Weiss behavior at high temperature, \( \frac{M}{H} =
\frac{C}{T-T_{CW}} \) , with $C$ the Curie constant and $T_{CW}$
the Curie-Weiss temperature.  The lower inset  in Figure \ref{MvsT
poly Xtal and film}(a) shows the magnetization derivative $dM/dT$
used to determine the transition temperature to the ordered state,
the Curie temperature, $T_C$. A Curie-Weiss fit of several
polycrystalline samples gives $T_{CW}$ = 155(6) K and $T_C$ = 98(2) K.
The effective magnetic moment, calculated from the Curie constant
is $p$ = 2.3(1) $\mu_B$. This value is lower but close to the one
expected for spin only Ni$^{2+}$, $p$ = 2.86 $\mu_B$.

A typical result for the single crystals is also shown in Figure
\ref{MvsT poly Xtal and film}(a), filled circles. In this case
$T_{CW}$ = 138(4) K and p = 2.60(4) $\mu_B$. We measured the single
crystals with the magnetic field applied along two crystallographic
directions, $H$ $\parallel$ (103) and $H$ $\parallel$ (010). In
the upper inset of Figure \ref{MvsT poly Xtal and film}(a) the susceptibility
extracted from magnetization measurements at 1 and 5 kOe ($\Delta
M/\Delta H = (M$(5 kOe)$-M$(1 kOe))/4 kOe) is shown. The inflection
point of these curves, identified with $T_C$, is located at 105 K
which agrees very well with the polycrystalline samples.

\begin{figure}[hhhhhhhhh]
\includegraphics[width=\columnwidth]{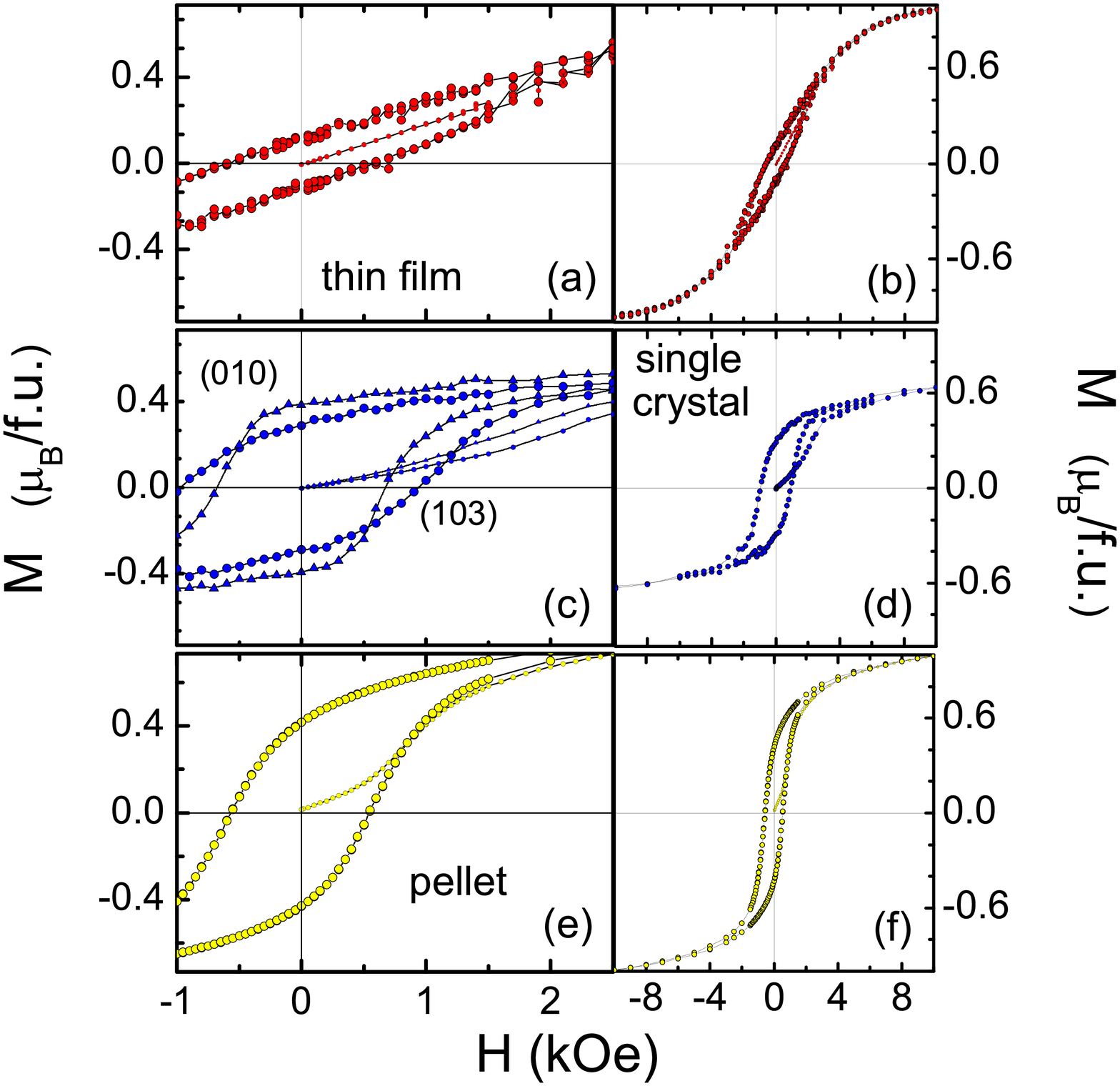} \caption[]{(color online)
M vs H at 2 K for a thin film (a) and (b), a single crystal (c) and (d) and
a polycrystalline pellet (e) and (f) of La$_{2}$Ni(Ni$_{1/3}$Sb$_{2/3}$)O$_6$.
For the single crystal(c), loops with the applied field along directions (103) and (010)
are included.}
\label{loopsraw}
\end{figure}

The thin films paramagnetic behavior ($T > T_c$) is very difficult
to isolate from the total signal, given their small mass,
typically about ~40 $\mu$g. However, subtracting the substrate
magnetization contribution, a clear transition to a ferromagnetic
state could be observed. Figure \ref{MvsT poly Xtal and film}(b)
shows the FC magnetization at 1 kOe, applied parallel to the
surface of the film ($H$ $\perp$ (001)). The low temperature
signal is approximately three times smaller than the one observed
on the polycrystals and single crystals. The inset in Figure
\ref{MvsT poly Xtal and film}(b) shows $dM/dT$, the peak indicates
a $T_C$ = 78 K, which is about 20 \% smaller than for bulk samples.

Hysteresis loops  ($M$ vs $H$) were measured at several
temperatures for $T < T_c$. Figure \ref{loopsraw} shows typical
loops at $T$ = 2 K for a thin film, a single crystal and a
polycrystalline pellet. For the crystals, loops with the applied
field along the (103) and (010) directions are included (Figure
\ref{loopsraw}(c)). In this case the initial magnetization branch,
the $M$ vs $H$ virgin curve measured after cooling at zero field,
falls always outside the loop area for temperatures bellow about
20 K. This effect can be interpreted as a sign of frustration, is
less visible in pellets and was not observed in films.

For temperatures below 20 K, we also measured the hysteresis loops after
cooling the sample in 1 T, observing that they coincide with those measured
after ZFC. No shift was found in the coercive fields ruling out
the presence of exchange bias \cite{exchange bias}.

\section{Discussion}

\subsection{The ordered state}
The value of the Curie-Weiss constant, the shape of the $M$ vs $H$
curves and the hysteretic behavior, all point to a ferromagnetic
state of the Ni$^{2+}$ below 100 K. However, the saturation
magnetization value, $M_s$, taken as the asymptotic extrapolation
with a Langevin function of the behavior at the largest applied
field (5 T), has a lower value  than the one expected for the
ferromagnetic complete polarization of the Ni$^{2+}$ magnetic moments,
2.67 $\mu_B$/f.u.. The experimental $M_s$ values range from 0.73
$\mu_B$/f.u. to 1.19 $\mu_B$/f.u. considering films, single
crystals and polycrystalline samples. These smaller experimental
values are better understood if the system behaves as a
ferrimagnet having two Ni$^{2+}$ magnetic sublattices
antiferromagnetically coupled, one at the $2d$ and another at the
$2c$ sites. The near 1/3 Ni$^{2+}$ random occupation of the $2c$
sites sublattice give as a result uncompensated Ni$^{2+}$ magnetic
moments that order at 100 K. For a perfectly stoichiometric sample
and full Ni$^{2+}$ occupancy of the $2d$ site $M_s$ should be 1.33
$\mu_B$/f.u., and lower values are expected if Sb$^{5+}$ partially
occupies also the $2d$ site. In particular, for the refined
occupancy of the octahedral sites in the powder, we can calculate
$M_s$ = 1.24 $\mu_B$/f.u., close to the measured values. A similar
ferrimagnetic order was reported in
Sr$_2$Fe(Fe$_{1/3}$Mo$_{2/3}$)O$_6$ \cite{CarbonioFeMo} and
Sr$_2$Fe(Fe$_{1/3}$U$_{2/3}$)O$_6$ \cite{CarbonioFeU} where the
magnetic Fe$^{3+}$ ions and non magnetic Mo$^{6+}$ or U$^{6+}$
display a similar structural arrangement and magnetic array as the
Ni$^{2+}$ and Sb$^{5+}$ in our case.

\begin{figure}[hhhhhhhhhhhh]
\includegraphics[width=.9\columnwidth]{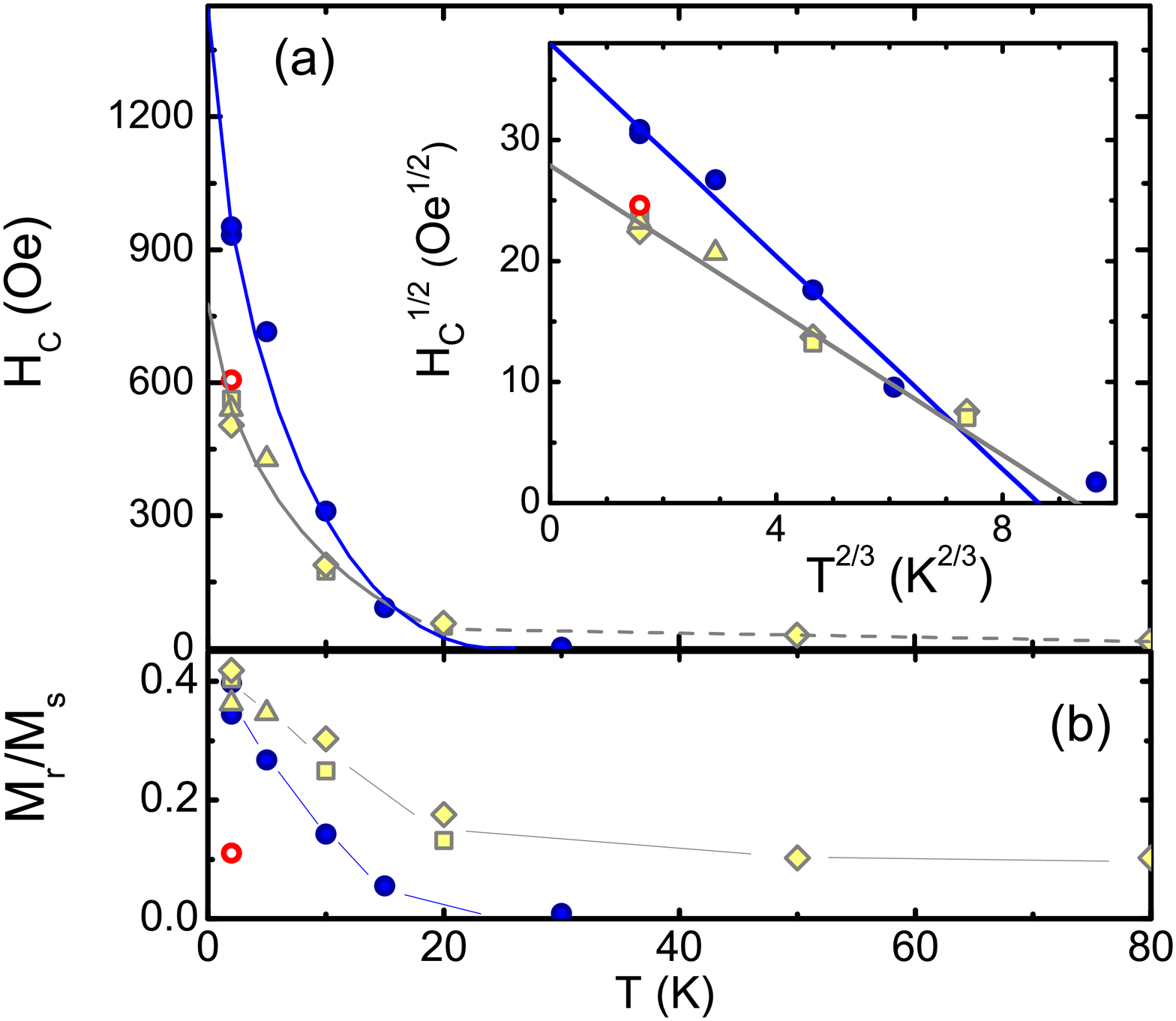} \caption[]{(color online)
(a) Coercive field $H_{c}$ vs $T$ for
La$_{2}$Ni(Ni$_{1/3}$Sb$_{2/3}$)O$_6$ polycrystalline pellets
(yellow, open symbols), single crystals (blue, dark circles) and
thin films (open circles). The lines corresponds to models
described in the text (SDWP and WDWP models, full lines and dashed
line respectively). Inset: H$_{c}^{1/2}$ vs T$^{2/3}$ showing the
linear behavior expected in the SDWP model. (b) Normalized
remanent magnetization (at $H$=0) vs $T$, the symbols represent
the same samples indicated in (a). The lines are a guide to the
eye.} \label{HcMvsT}
\end{figure}

In order to get a microscopic understanding of the uncommon
features seen in the virgin curves we shall analyze with some
detail the characteristics of the ordered state through a study of
the irreversible magnetization measurements from hysteresis loops.
Figure \ref{HcMvsT} (a) shows that the coercive field for the
single crystals (measured along the (103) direction) has a similar
temperature behavior as that of the polycrystalline pellet at low
temperatures \cite{IEEE Arxiv}. We have previously established
that the polycrystalline material coercive field has an upturn at
$T\!\approx$ 20 K changing from a weak domain wall pinning (WDWP)
behavior at high temperatures to a strong domain wall pinning
(SDWP) one below 20 K \cite{IEEE Arxiv}. Analyzing the coercive
field and the time dependence of the magnetization we have ruled
out the freezing of large particles as a posible mechanism for the
$H_c$ upturn \cite{IEEE Arxiv}. We found that in the single
crystals the coercivity above 30 K is negligibly small. The film
coercivity, measured at 2 K, is close to the bulk value, in spite
of the expected $H_c$ enhancement due to barriers for the DW
motion introduced by surface roughness or local strains
\cite{Steren}. Figure \ref{HcMvsT}(b) shows the $T$ dependences of the
the ratio between remanent and saturation magnetizations
$(M_r/M_s)$. This ratio and $H_c$ increase steeply when the
temperature is lowered below 20 K indicating increase in the
energy needed to change the direction of $M$.

In the low temperature regime, $T\lesssim$ 20 K, the
polycrystalline pellets and single crystals ($H$ $\parallel$ 103)
$H_c$ data can be described by a model of strong domain wall pinning
(SDWP),
\begin{equation} \label{HcSDWP}
H_{c} = H_{0S} \left[ 1-\left( \frac{75 k_B T}{4bf} \right)^{2/3}
\right] ^2
\end{equation}
\noindent where $H_{0S}$ is the coercive field at zero temperature,
$f$ is the magnetic force needed to depin a domain wall and
$b$ is a measure of the domain wall thickness. The fitted values
are shown in Table \ref{HcSvfits}.

An estimation of $b$ from the exchange stiffness, $A$, and
anisotropy constant, $K_1$, yield a small value of the domain wall
thickness \( b=\pi(\frac{2A}{K_1})^{1/2}\backsimeq 10 nm \). The
anisotropy constant was calculated at 2 K from the area between
the anhysteretic curves $M$ vs $H$  and the $M$ axis
\cite{Cullity} for the single crystals measured in the (103) and
(010) direction. The value obtained was $K_1$= 3 10$^{5}$
erg/cm$^3$ \footnote{This value was obtained considering the 010
direction as the easy magnetization axis}.
 Although the calculated anisotropy constant is of the order
of the value found in other ferrimagnets \cite{Cullity}, the
exchange stiffness constant is small due to the rather long
average distance between uncompensated Ni$^{2+}$ moments in the
structure.

\begin{table}
\caption{\label{HcSvfits}Fitted parameters for the SDWP model
below 20 K, for polycrystalline, $PC$, and single crystalline,
$SC$, samples.}
\centering  \scalebox{1}{
\begin{tabular}[b]{l c c c c c c }
\hline \hline
&     &   $4bf$    &   $H_{0S}$   \\
&    &    (10$^{-13}$erg)   &   (Oe)   \\
\hline \hline
 $PC$   &   &   3.07    &   780    \\
$SC$   &   &   2.54    &   1440   \\
  \hline \hline
\end{tabular}}
\end{table}

The microscopic origin of the change of regime of domain wall
pinning mechanism at 20 K remains
unclear. The onset of a disordered or frustrated magnetic state
may result in the emergence of strong pining sites for DW
movement. The answer could be made evident analyzing the virgin
curve in the hysteresis loops. The initial branch of the $M$ vs
$H$ curve (cooling from above $T_c$ at zero applied field) does
not fall entirely within the magnetization loop for all the
samples, except for the thin film, see Figure \ref{loopsraw}. In
what follows this uncommon behavior will be addressed.

\begin{figure}[hhhhhhhhhhhhh]
\includegraphics[width=0.9\columnwidth]{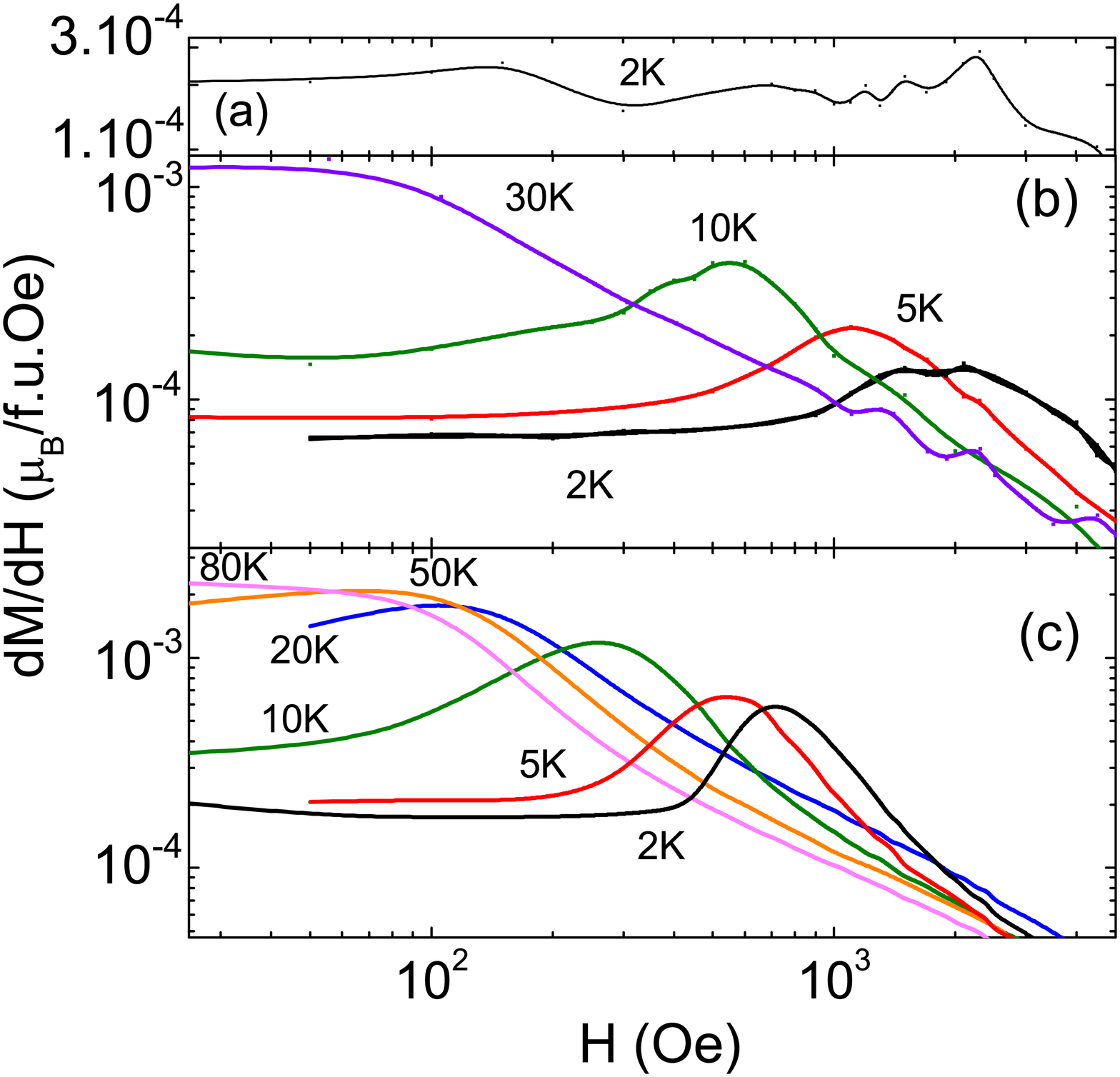} \caption[]{(color online)
Derivative of the virgin curves in Magnetization loops for (a)
thin film, (b) single crystals and (c) polycrystalline pellets.}
\label{Rinis}
\end{figure}

\subsection{The virgin magnetization curve}

Similar loops as those shown in Figure \ref{loopsraw} were
obtained for several temperatures, below and above the temperature
of the steep increase of the coercive field when lowering $T$. We
have found that the virgin curve excursion outside the regular
loop takes place at low temperatures ($T \lesssim$ 20 K) coinciding
with the regime of SDWP described previously.

In some systems the virgin curve was observed to go outside the
loop for a certain range of fields and temperatures
\cite{Alejandro, Morales, Joy, Sarkar, Zysler, Senoussi,
Patankar}. In some complex magnetic oxides this feature was
related with magnetic cation disorder \cite{Alejandro, Morales},
which led not only to a spin-glass behavior but also to local
structural distortions. These local distortions are due to a
microscopic rearrangement of valence electrons \cite{Joy}, or to
magnetic cations deficiency changing the nature of the local
crystal field \cite{Sarkar} which in turn lead to an
irreversible movement of domain walls \cite{Joy, Sarkar}. The
anomalous virgin curve in a ferrimagnet was also attributed to the
development of an antiferromagnetic order at low temperature that
produces a magnetic glass state \cite{Patankar}.

\begin{figure}[hhhhhhhhhhhhhhh]
\includegraphics[width=0.95\columnwidth]{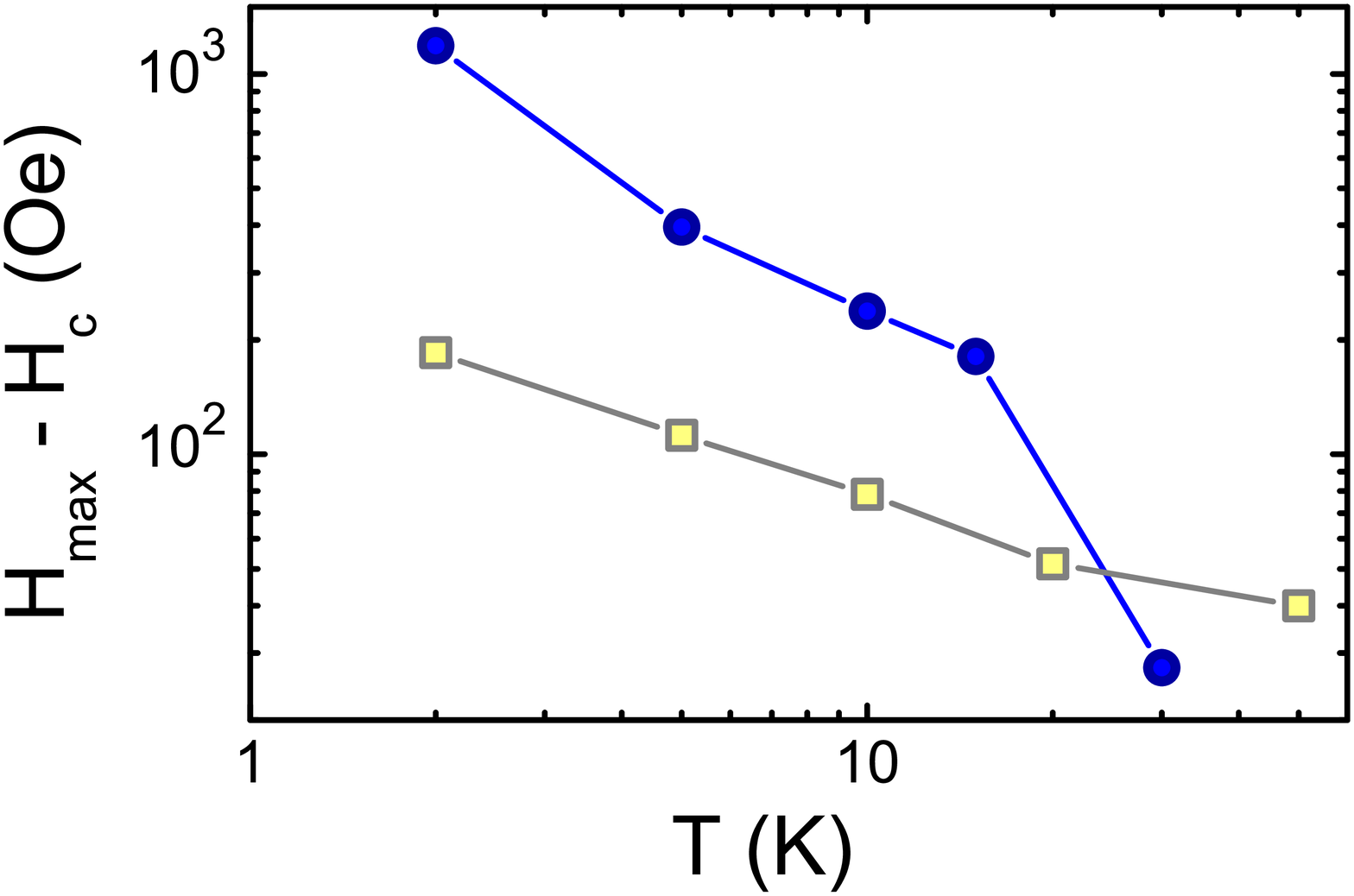} \caption[]{(color online)
Difference of the field of the maximum slope of virgin curves in
magnetization loops and the coercive field for crystals (blue,
dark circles) and polycrystalline pellets (yellow, open squares).
The lines are guide to the eye.} \label{deltaHvsT}
\end{figure}

Figure \ref{Rinis} shows the $H$ derivative of the virgin curves
dM/dH at several temperatures for the thin film, single crystals and
polycrystalline pellets. In the thin film case  (Figure
\ref{Rinis}(a)), the derivative is approximately constant. For all
the other samples there is a maximum in the derivative that
indicates a characteristic field, $H_{max}$, for the magnetic
moments alignment, larger than the coercive field at low $T$
(Figure \ref{deltaHvsT}).

In a spin glass scenario $H_{max}(T)$ could represent the line
separating the glassy phase from the ordered state.
Therefore a $T$ dependence following the
Almeida-Thouless \cite{Almeida} or Gabay-Tolouse \cite{Gabay} models
could be expected.  This is not our case, indicating a more
complex behavior involving the hindering of DW movement. We have
also ruled out the magnetic glass \cite{Roy} scenario because we
found no evidence for a long range antiferromagnetic order at low
temperatures and we did not detect differences between FC cooling
and FC warming magnetization measurements at 1 kOe as in a typical
magnetic glass displaying the kinetic arrest of the phase
transition \cite{SarkarII}. The difference between $H_{max}$
and $H_c$ (Figure \ref{deltaHvsT}) is larger for single crystals
than for polycrystalline pellets suggesting an intrinsic origin
for this behavior.

\subsection{The frustrated state and its removal}

To understand the microscopic origin of the frustration the local
magnetic interactions of Ni$^{2+}$ have to be analyzed.  They
depend on the nearest magnetic neighbor (1 nnn) and on the exchange
path for the second and third nearest neighbor (2 nnn, 3 nnn).
Considering the structure (Figure \ref{structure}(left)) with
perfect stoichiometry and full Ni$^{2+}$ occupancy of the $2d$
site, the interaction among 1 nnn relays on the presence of a
Ni$^{2+}$ in the closer $2c$ site (1/3 Ni$^{2+}$ ion occupancy).
If this is the case there will be an antiferromagnetic
superexchange interaction of the Ni$^{2+}$-O-Ni$^{2+}$ kind
\cite{Goodenough}. The remaining 2/3 of the $2c$ sublattice is
occupied by  non magnetic (d$^{10}$) Sb$^{5+}$ ions. Then, a
Ni$^{2+}$ ferrimagnetic lattice is formed at the $2d$ and $2c$
sites. The $2d$ site Ni$^{2+}$ second and third nearest neighbor
interactions are mediated by -O-O- (at $\sim$ 90$^{o}$) and
-O-Sb$^{+5}$-O- (at $\sim$ 180$^{o}$) super-superexchange paths
(see Figure \ref{structure}(right)), whose relative strength would
determine the type of magnetic ordering in the regions with
Sb$^{+5}$ in the $2c$ sublattice. These super-superexchange
interactions were found to be dominant for the antiferromagnetic
structure of some ordered double perovskites A$_2$BB'O$_6$ where B
is a magnetic transition metal ion and B' is a non magnetic cation
\cite{Rodriguez}. In particular, LaSrNiSbO$_6$ was found to have
an antiferromagnetic structure of type I (ferromagnetic planes
parallel to the $ab$ and $ac$ planes antiferromagnetically coupled
along the $c$ and $b$ direction respectively) with a transition
temperature of 26 K \cite{Attfield}. Therefore, for
La$_{2}$Ni(Ni$_{1/3}$Sb$_{2/3}$)O$_6$, at temperatures around 20
K, the Ni$^{2+}_{2d}$-Ni$^{2+}_{2d}$ interactions mediated through
-O-O- paths and the ones mediated through -O-Sb$^{5+}$-O-, both
antiferromagnetic in nature, became important. These interactions
 together with Ni$^{2+}$-O-Ni$^{2+}$ superexchange one that exists
 below 100 K, create a magnetically frustrated state at lower temperatures
 than 20 K. A similar antiferromagnetic transition temperature was
 found in related Co based perovskites \cite{Fuertes} and in
La$_{2}$Ni(Ni$_{1/3}$Nb$_{2/3}$)O$_6$ \footnote{
La$_{2}$Ni(Ni$_{1/3}$Nb$_{2/3})$O$_6$ shows an antiferromagnetic
ordering at 28 K and is currently under study.}.
\begin{figure}[hhhhhhhhhhh]
\includegraphics[width=.9\columnwidth]{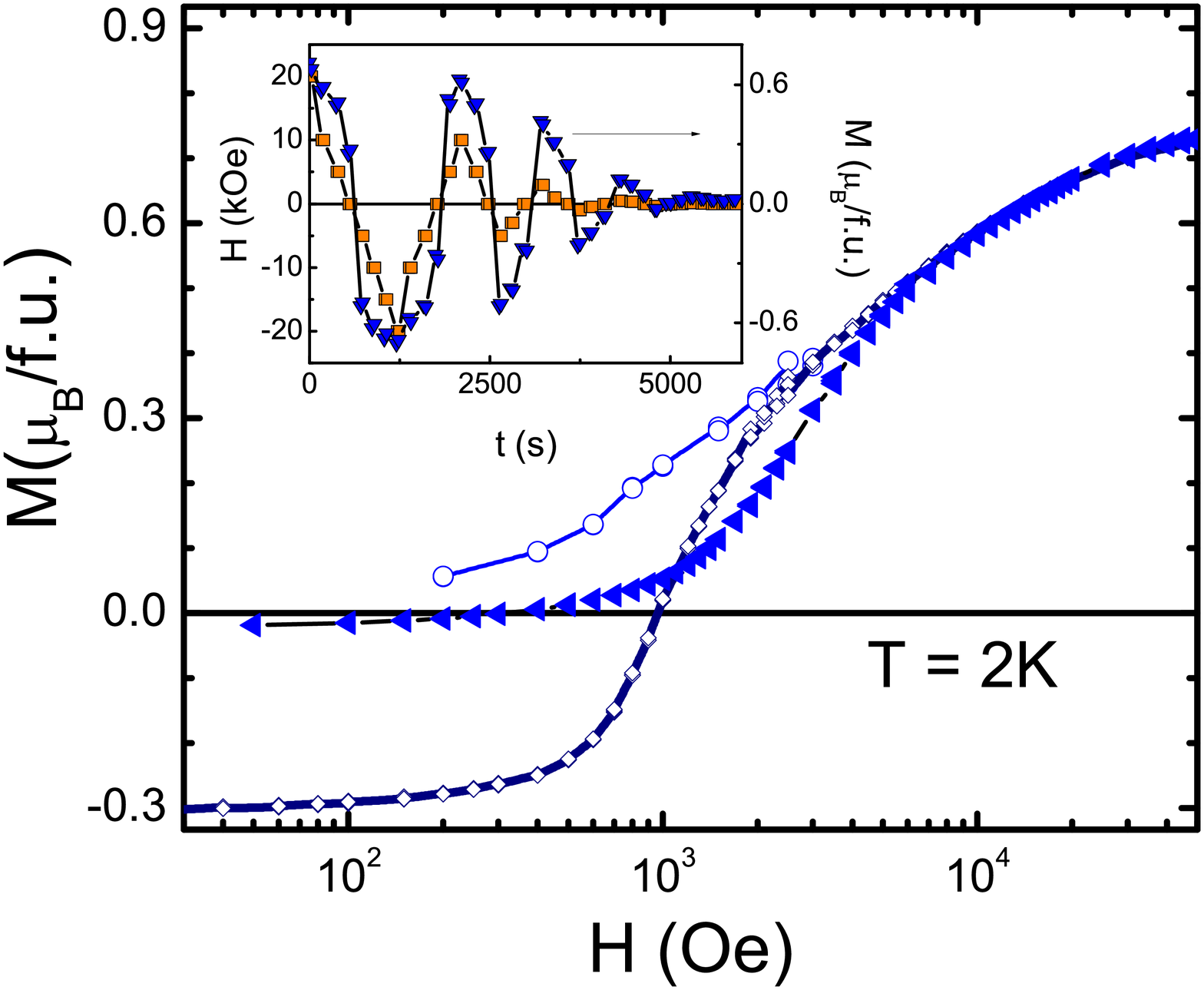} \caption[]{(color online)
$M$ vs $H$ for a single crystalline sample at 2K showing the
initial branch after a demagnetizing process (circles). The virgin
curve (triangles) and the regular loop ascending field branch
(diamonds) are also shown. Inset: $M$ and $H$ vs time in a
demagnetizing process.} \label{demagnRaw}
\end{figure}
Consequently, a magnetically frustrated ground state is expected
when cooling the sample in zero field due to competing
antiferromagnetic interactions among the Ni$^{2+}$ ions and
$H_{max}$ is necessary to overcome this frustration. The
microscopic origin of the barrier which is overcome by increasing
$H$ at low $T$ is still not clear. It is probably related to the
spin orientation of the Ni$^{2+}$ moments interacting
antiferromagneticaly with its 1 nnn trying to keep their
antiferromagnetic arrangement in planes perpendicular to $H$ and
the uncompensated Ni$^{2+}$ moments following the field as in a
normal ferromagnet. The random orientation of the
antiferromagnetic regions and the frustrated interaction near
Sb$^{5+}$ ions hinders the initial movement of the DW. When they
are oriented, further changes of the magnitude of $H$  result only
in canting of the moments. However, the canting is not very
important since a magnetization saturation value seems to be
almost achieved at 5 T with $M_s$ corresponding to the
uncompensated Ni$^{2+}$ moments in the ferrimagnetic structure, as
described in previous sections. This model can be tested going
back to the zero magnetization state with a demagnetizing
protocol, illustrated in the insets of Figure \ref{demagnRaw}.
After this demagnetizing procedure, the zero magnetization state
would involve only a random orientation of the uncompensated
Ni$^{2+}$ moments and the antiferromagnetically arranged regions
would have the moments perpendicular to the direction of the
preexisting field, providing a new ground state. Indeed, the
results indicate that after this demagnetizing process, the sample
initial magnetization branch lies inside the hysteresis loop. This
is illustrated in Figure \ref{demagnRaw} for a single crystal at
$T$ = 2 K, where the virgin magnetization, the regular loop branch
and a new initial magnetization branch are shown. This new initial
magnetization is measured after the demagnetizing protocol shown
in the inset.

The absence of exchange bias \cite{exchange bias} in the
magnetization loops indicate that large clusters with only
antiferromagnetic super exchange Ni$^{2+}$-O-Ni$^{2+}$
interactions are not likely to exist (i.e: the one third Ni$^{2+}$
occupancy of the $2c$ site seems to be homogeneous at a
microscopic level).

\begin{figure}[hhhhhhh]
\includegraphics[width=\columnwidth]{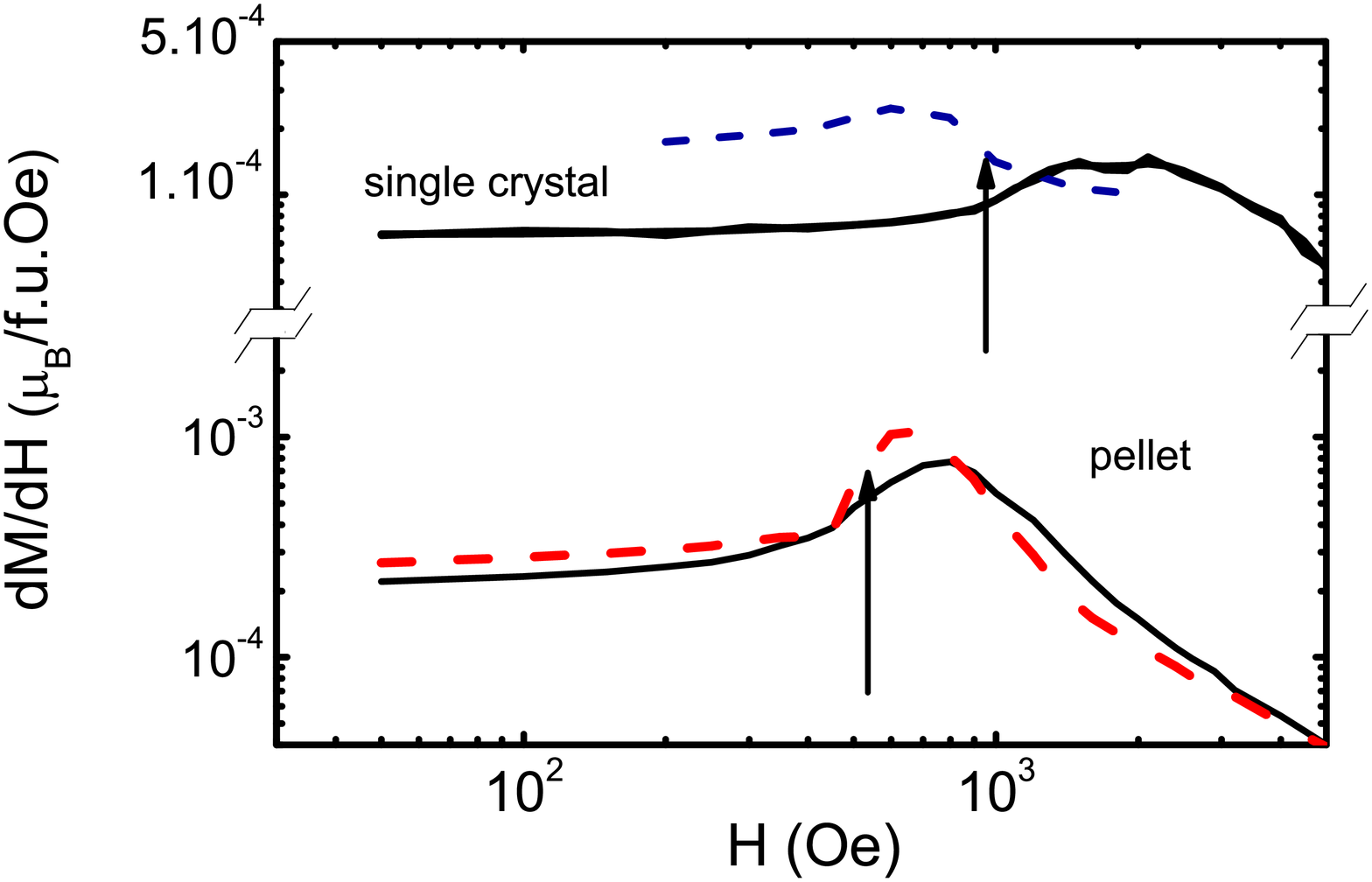} \caption[]{(color online)
Comparison of the derivatives of the virgin curves in
magnetization loops and the initial magnetization after a
demagnetizing process, for a single crystal(a), a polycrystalline
pellet (b). All the curves were taken at 2K. The arrows indicate
the corresponding coercive field} \label{RinisDemag}
\end{figure}

Figure \ref{RinisDemag} shows the comparison between the
derivatives of the virgin curves and the initial magnetization
after a demagnetizing protocol at 2 K for a single crystal and  a
polycrystalline pellet. We performed similar experiments at 5 and
10 K. In all the cases the new initial magnetization lies inside
the loop as in a usual, non- frustrated, ferromagnet.

We can observe in Figure \ref{RinisDemag} that the initial
susceptibility of the virgin curve after isothermal
demagnetization is larger than the corresponding to the thermally
demagnetized sample indicating that the new ground state is easier to
magnetize in the direction of the preexisting magnetic field.

For the thin films, no indication of the frustrated state was
found (Figures \ref{loopsraw}(a) and \ref{Rinis}(a)). The
superexchange and super-superexchange interactions are very
sensitive to bond angles and lengths \cite{Hoffman}, which are
likely modified near the surface. In our samples, the magnetic
anisotropy imposed by the geometry seems to overcome the
disordered interactions that sets in at low temperatures.

\section{Conclusions}

It is shown that the anomalous behavior of the virgin curve in
hysteresis loops is a distinctive feature of a low temperature
magnetic frustrated state found in the ferrimagnetic double
perovskite oxide La$_{2}$Ni(Ni$_{1/3}$Sb$_{2/3}$)O$_6$. The virgin
curve lies outside the loops at $T$ $\lesssim$ $20$ K,
at about one fifth of the ferrimagnetic ordering temperature
($T_{c}\!\approx $ 100 K). This was found to be an indication of a
microscopically irreversible process possibly involving the
interplay of antiferromagnetic interactions that hinder the
initial movement of domain walls.  This feature was observed in
single crystals and in polycrystalline pelletized samples but not
in thin films. This initial frustrated magnetic state is overcome
by applying a characteristic field. Above this field the material
behaves macroscopically  as a typical ferromagnet. The model
proposed for the frustrated state is based on the competing
antiferromagnetic interaction between 1 nnn and 3 nnn (ie:
Ni$^{2+}$-O-Ni$^{2+}$ and Ni$^{2+}$-O-Sb$^{5+}$-O-Ni$^{2+}$). This
microscopic scenario for the frustrated state is tested by going
back to the zero magnetization state at fixed low temperature
applying a demagnetizing protocol.

\section{Acknowledgments}

We thank P. Pedrazzini for help with the crystal growth and E. De
Biasi for fruitful suggestions. R.E.C., E.E.K., and G.N. are
members of CONICET, Argentina. D.G.F. has a scholarship from
CONICET, Argentina. Work partially supported by ANPCyT
PICT07-00819, CONICET PIP11220090100448 and SeCTyP-UNCuyo 06/C381.
R.E.C. thanks FONCYT (PICT2007 00303), CONICET (PIP
11220090100995) and SECYT-UNC (Res. 214/10) for finantial support.
E.E.K. thanks ANPCyT PICT2008-1731 for finantial support.


\end{document}